\documentclass{mn2e}
\usepackage{epsfig}
\usepackage{pslatex}
\usepackage{amssymb}

\newcommand{\ltsimeq}{\raisebox{-0.6ex}{$\,\stackrel
        {\raisebox{-.2ex}{$\textstyle <$}}{\sim}\,$}}

\def\msun{{\rm M_{\odot}}}
\def\rsun{{\rm R_{\odot}}}

\def\today{\number\year \ \ifcase\month\or
  January\or February\or March\or April\or May\or June\or
  July\or August\or September\or October\or November\or December
 \fi \ \number\day }
\date{Accepted ??. Received ??; in original form \today}

\volume{000}

\pagerange{\pageref{firstpage}--\pageref{lastpage}} \pubyear{2004}

\begin{document}

\label{firstpage}

\title[On the metallicity dependence of HMXBs]
{On the metallicity dependence of HMXBs}

\author[L.\,M. ~Dray]{L.\,M. ~Dray\thanks{E-mail: Lynnette.Dray@astro.le.ac.uk}\\  
Theoretical Astrophysics Group, University of Leicester,
Leicester, LE1~7RH, UK\\
}

\maketitle

\begin{abstract}
It is commonly assumed that high mass X-ray binary (HMXB) populations 
are little-affected by metallicity. However, the massive stars making up 
their progenitor systems depend on metallicity in a number of ways, not 
least through their winds. We present simulations, well-matched to the observed 
sample of Galactic HMXBs, which demonstrate that both the number and the mean period
of HMXB progenitors can vary with metallicity, with the number increasing by about a factor 
of three between solar and SMC metallicity. However, the SMC population itself cannot 
be explained simply by metallicity effects; it requires both that the HMXBs observed 
therein primarily sample the older end of the HMXB population, and that the star 
formation rate at the time of their formation was very large.

\end{abstract}


\section{Introduction}
The high mass X-ray binaries represent a late stage in the life of a close massive binary, when 
one star has collapsed to a neutron star (NS) or black hole (BH). Accretion of matter lost in a wind or by
Roche lobe overflow from its companion onto the compact object results in X-ray emission. As distinct from the 
low mass X-ray binaries (LMXBs), which may become luminous considerably later than the formation of the compact 
object and after considerable orbital evolution (e.g. Verbunt \& ven den Heuvel 1995), the HMXB phase is 
though to occur relatively soon after the first SN in a system, and cannot last longer than the nuclear 
burning time ($ \ltsimeq 3 \times 10^{7} {\rm yr}$) of the massive companion star. The occurence of HMXBs is 
therefore linked to recent star formation in a similar manner to massive stars.

The population is itself split into two subgroups, depending on the spectral type of the companion star 
(van Paradijs 1983). Approximately three-quarters of observed Galactic HMXBs have Be star companions. These systems 
have long periods and eccentric orbits; X-ray emission is often observed at periastron, when the NS interacts with the 
disk around the Be star. In contrast, the rest of the HMXB population is made up of OB supergiant systems, for which 
usually persistent emission is powered by the strong wind of the companion star prompting disk formation around the NS, 
or very occasionally by Roche lobe overflow (RLOF). These are generally shorter-period systems which have low eccentricity, 
often because the orbits have circularized. OB supergiant systems have significant runaway velocities whereas Be systems 
do not (Chevalier \& Ilovaisky 1998), which suggests different evolutionary paths are involved (van den Heuvel et al. 
2000). Due to the transient nature of the Be/X-ray binaries, it is likely that the true fraction of the population they 
represent is in fact significantly larger than that observed. 

The question of whether HMXBs display metallicity dependence is an interesting one. It is commonly assumed that 
binary evolution is little-affected by the metallicity of the stars involved. However, this assumption may break 
down in the case of massive stars, since the loss of mass and angular momentum through winds during their 
lifetimes is non-negligible. A massive star which goes through a Wolf-Rayet phase may lose over half its mass in winds 
even if there is no binary interaction (e.g. Maeder \& Meynet 2000). The exact metallicity dependence of massive star 
winds is still a matter of debate (e.g. Vink \& de Koter 2005) but, since the winds of massive hot stars are 
line-driven, decreasing the metallicity is likely to lower the mass-loss rate. A ${\rm Z^{0.5 - 0.7}}$ dependence 
for mass loss has been suggested (Kudritzki et al. 1989; Vink, de Koter \& Lamers 2001). This may affect both the 
evolutionary path the system takes and also the compact object mass, both of which in turn may affect whether the 
system is split by the first supernova or goes on to undergo a HMXB phase. Theoretically, lower mass loss
should mean greater rotation rates for lower-metallicity massive stars, since more angular momentum is retained. If 
this persists until the HMXB stage, it may increase the incidence of the propeller effect (Shtykovskiy \& Gilfanov 2005a)
which impedes accretion onto the NS. However, it should be noted that observationally there is some evidence that 
metallicity does not have a strong effect on rotation rates (Penny et al. 2004). A star of lower metallicity 
will also generally have a smaller radius, which in turn can delay the onset of RLOF with respect to a 
comparable high-metallicity system. 

The population synthesis of binary black holes carried out by Belczynski, Sadowski 
\& Rasio (2004, see also Belczynski et al. 2004b) finds an increased binary BH population at low Z, 
which they attribute both to lower mass-loss 
rates leading to a greater BH population and also to the greater retention of angular momentum leaving 
binaries harder to disrupt. This contrasts with the LMXB population, which may show an increase at higher 
metallicity (Maccarone, Kundu \& Zepf 2004). Hurley, Tout and Pols (2002) also find an increase in HMXB population 
between $Z = 0.02$ and $Z=0.0001$.
There are a large number of population synthesis calculations for or involving HMXBs 
carried out at solar metallicity: see e.g. Podsiadlowski et al. (2004) and references therein;
Sepinsky, Kalogera \& Belczynski (2005) and references therein; 
Dalton \& Sarazin (1995); Pfahl et al. (2002); Portegies Zwart \& Verbunt (1996); 
Brandt \& Podsiadlowski (1995); Van Bever \& Vanbeveren (2000); Kalogera (1996).

Observationally, it is hard to pick out trends which cannot be explained by other means. The Small Magellanic Cloud
(SMC), with a metallicity of ${\rm ~ 0.2 Z_{\odot}}$, appears overabundant in HMXBs by a factor of about 50 per unit stellar mass 
when compared to the Milky Way (Majid, Lamb \& Macomb 2004). It is likely, however, that this is primarily due to recent 
intense star formation. It is notable amongst the large SMC HMXB population that only one source (SMC X-1) has an OB 
supergiant companion. For the Large Magellanic Cloud (LMC), which is intermediate in metallicity between the 
SMC and the Milky Way, the 
HMXB population appears similar to the Galactic one (Neguerela \& Coe 2002) albeit with a deficit of low-luminosity 
sources which may be due to the propeller effect (Shtykovskiy \& Gilfanov 2005a). 

To a first approximation, the X-ray luminosity function of HMXBs obeys a universal power law distribution; it has 
been suggested that X-ray luminosity would therefore be a possible star formation rate (SFR) indicator in distant galaxies 
(Grimm et al. 2003). This requires the number of HMXBs produced by a given amount of star formation to be roughly 
constant with metallicity. 

We have previously carried out simulations aimed at reproducing the populations of runaway O and WR stars in the 
Galaxy and LMC (Dray et al. 2005, hereinafter D05). Over half of runaways get their high velocities from supernova 
explosions in binary systems (Hoogerwerf, de Bruijne \& de Zeeuw 2001), most of which lead to the splitting of the 
binary system. It is an inevitable result of such simulations that one also obtains the parameters of a simulated 
population of binaries which are not separated. These are the HMXB progenitors. In this paper we consider their 
match to observed HMXB systems and the metallicity trends they reveal.

\section{Simulations}
Details of the Monte Carlo simulations carried out are given in D05. The main ingredients are as follows:
\begin{itemize}
\item It is assumed that at formation massive stars are nearly all in binaries (Mason et al. 1998) and that the
initial distribution of binary parameters is flat in $q = M_{2}/M_{1}$ and $\log P$. For 
ease of comparison between metallicities we also use a constant star formation rate here unless 
otherwise specified. In order to sample the HMXB population well we run larger numbers of systems than D05 -- 
up to 15 million initial binaries at each metallicity.
\item Binary evolution until contact or the second SN is taken from the grid of  
simulations of Dray \& Tout (2005) using the Eggleton code (Eggleton 1971, Pols et al. 1998 
and references therein), which 
cover metallicity from $0.02$ to $0.001$ and conservative and non-conservative evolution. They include 
a full treatment of massive star mass loss, accretion and thermohaline mixing. We also run simulations 
using simple analytic assumptions for stellar lifetimes and masses to check that the results from the full 
evolutionary models are reasonable.
\item In the case of non-conservative evolution we assume that, for any given episode of mass transfer, 
only the first ten percent of transferred matter is accreted (see e.g. Packet 1981). The rest is lost from the system. For angular 
momentum loss we use the prescription of Hurley et al. (2002).
\item For common-envelope systems, we use the prescription of Webbink (1984) as stated by Dewi \& Tauris (2000), 
with $\eta_{\rm CE} = 1.0$ and $\lambda = 0.5$. Other reasonable values of these parameters do not significantly 
change the results. Merger systems are assumed to evolve similarly to secondary stars which have accreted a similar amount of mass.
\item The amount of mass lost in the SN is calculated using the fit to mass loss with core mass from the models 
of Woosley \& Weaver (1995) by Portinari, Chiosi \& Bressan (1998). It is assumed that SN kicks are isotropic and 
follow a Maxwellian distribution with mean $450 {\rm km\,s^{-1}}$ (Lyne \& Lorimer 1994; note that more recent studies 
(Hobbs et al. 2005) find similar results) and that only stars which undergo 
direct collapse to a black hole (here taken to be those stars with final core masses over $15 \msun$) do 
not experience kicks. Restricting the kick velocities imparted to e.g. close systems (Pfahl et al. 2001) 
or BH-forming explosions 
results in difficulty fitting the observed runaway populations, but a slightly smaller mean kick value (e.g. 
$300 {\rm km\,s^{-1}}$) does not substantially change the qualitative results. In the current paper it is 
assumed that the maximum mass of a neutron star is $2.2 \msun$ (Akmal, Pandhariprinde \& Ravenhall 1998); 
it should be noted that this value is poorly-known, depending strongly on the 
assumed equation of state (Cook, Shapiro \& Teukolsky 1994) and that its value may vary between stars depending 
on their rotation rates (Akiyama \& Wheeler 2005).
\item The effect of the kick on the binary parameters is taken from Tauris \& Takens (1998). For quantities not 
given in that paper, we use equations from Brandt \& Podsiadlowski (1995).
\end{itemize}
As can be seen in D05, these ingredients reproduce the Galactic WR and O star runaway populations well, and 
are reasonably stable to changes in the input parameters. Preliminary investigations in D05 also showed 
that binaries which survived the first SN reproduced well the runaway velocity distribution of HMXBs, 
with short-period systems able to attain higher runaway velocities. 

\begin{table*}
\begin{minipage}{\textwidth}
  \caption{Masses of some HMXB components in the literature}
  \begin{tabular}{@{}lrrrlr@{}}
 \bf & \bf ${\rm\bf M_{compact}/\msun}$\footnote{Note that full orbital solutions are more 
commonly obtained for bright, short-period systems, so this population is dominated by selection effects.}
& ${\rm\bf M_{OB}/\msun}$ & \bf period/days & \bf reference& 
\\ 
\hline
Vela X-1\footnote{The first set of mass values for Vela X-1 are obtained assuming the companion star is Roche lobe-filling; the second set of values assume an inclination angle of $i = 90\deg$.}& $2.27\pm0.17$ & $27.9\pm1.3$ & 8.96 & {\small Quaintrell et al. 2003}&\\ 
         & $1.88\pm0.17$ & $23.1\pm1.3$ & 8.96 & {\small Quaintrell et al. 2003}&\\ 
Cen X-3  & $1.21\pm0.21$ & $20.5\pm0.7$ & 2.09 & {\small Ash et al. 1999}&\\ 
1538-522 (QV Nor)  & $1.3\pm0.2$ & $19.8\pm3.3$ & 3.73 & {\small Reynolds, Bell \& Hilditch 1992}&\\ 
4U 1700-37 & $2.44\pm0.27$ & $58\pm11$ & 3.41 & {\small Clark et al. 2002}&\\ 
SS433    & $\sim 9$\footnote{Error bars for plotting have been estimated from alternative model fits in the source paper for these values, since no error estimates are given.}
      & $\sim 30$ & 13.1 & {\small Cherepashchuk et al. 2005}\\
Cyg X-1  & $10.1^{+4.6}_{-5.3}$ & $17.8^{+1.4}_{-6.1}$ & 5 & {\small Herrero et al. 1995}\\
LS 5039  & $3.7^{+1.3}_{-1.0}$ & $22.9^{+3.4}_{-2.9}$ & 3.91 & {\small Casares et al. 20055}\\
LMC X-1  & $4^{+0.5}_{-0.2}$ & $\sim 20$ & 4.22 & {\small Yao, Wang \& Zhang 2005; Hutchings et al. 1987}\\
LMC X-3  & $4.19\pm0.02$ & $1 - 3?$ & 1.70  & {\small Yao, Wang \& Zhang 2005; Kuiper, van Paradijs \& van der Klis 1988}\\
LMC X-4  & $1.34^{+0.48}_{-0.44}$ & $14.6^{+2.4}_{-1.8}$ & 1.4 & {\small Pietsch et al. 1985}\\
SMC X-1  & $1.05\pm0.09$ & $15.5\pm1.5$ & 3.89 & {\small Van der Meer et al. 2005}\\
PSR J0045-7139\footnote{This SMC system is not a HMXB, but consists of a NS and a massive star which may go through a HMXB phase.}  & $1.58\pm0.34$ & $10\pm1$ & 51.17 & {\small Thorsett \& Chakrabarty 1999}\\
\hline
\end{tabular}
\end{minipage}
\end{table*}

\section{Post-SN evolution and accretion}

The distribution of system parameters from the simulations detailed in the previous section represent 
potential HMXB {\it progenitors} rather than HMXBs themselves. For an X-ray binary to be bright enough to be observed, 
sufficient accretion onto the compact object must take place. This requires at the least a strong wind from the companion and 
that the stars are relatively close at some point in their orbit. Even if 
these conditions are fulfilled, other circumstances may prevent observable accretion occurring.
In the following analysis, we neglect long-period direct collapse systems, since predominantly they have 
low-eccentricity orbits and remain at large separations throughout their 
lifetimes\footnote{It should be noted that, whilst Mirabel \& Rodrigues (2003) claim that 
Cyg X-3 contains a BH formed by direct collapse because its motion is not significantly different from its 
parent association Cyg OB3, for an isotropic distribution of SN kicks it is perfectly possible to obtain 
a small number of systems which are only imparted very small velocities by the SN because the kick may partially cancel out the velocity 
effect of off-centre mass loss (see e.g. the binary velocity distributions shown in D05).}.

For other systems we must consider a number of factors which influence whether the systems will be X-ray bright or 
not and what their observable characteristics are likely to be. In order to assess the luminosity and parameter 
evolution of our sample of systems, we use again the large library of binary stellar tracks utilised in D05. 
For each system which survives until one component undergoes a SN explosion we have a track corresponding 
to the evolution of the secondary after the explosion of the primary, assuming no further interaction. Due to the enormous 
parameter space imposed by the possible variation in SN kick size and direction for each individual system, it is not 
practical to run full evolutionary simulations for each post-SN orbital outcome. Rather we use the no-interaction tracks to 
calculate the orbital evolution for non-interacting and wind-accreting systems, and apply synthetic prescriptions for 
orbital evolution in the case of RLOF. 
 
For systems which undergo common envelope (CE) evolution the full evolutionary tracks end at the onset of 
this phase. However, comparison of the non-CE tracks we have suggests that the main factors governing the evolution
of the secondary after the SN of the primary are its original mass, the amount it has accreted and, to a lesser
extent, the age at which that accretion occurred. Since we have post-primary-SN evolutionary tracks for three 
different levels of mass transfer, this provides a relatively well-populated grid in initial mass, mass accreted and 
time of mass transfer. We then represent post-CE post-SN secondaries by the closest grid model in these quantities,
 assuming no further accretion onto the secondary during the CE phase. For stars which accrete only small amounts of 
matter ($\ltsimeq 0.1 \msun$) before the onset of the common envelope phase, we use a single star model of the appropriate 
mass and age instead.

Once we have a suitable post-SN evolutionary track, we use it to calculate the orbital evolution, wind accretion and 
lifetimes for the binary parameters of the system the star is in. The X-ray luminosity of a system is governed by the 
accretion rate onto the compact object. For wind-accreting systems this is usually taken to the the Bondi-Hoyle rate 
(Bondi \& Hoyle 1944). We use the formulation of this rate and of the corresponding angular momentum accretion as 
given in Hurley et al. (2002), with 

\begin{equation}
\langle \dot{M}_{\rm 2A} \rangle = \frac{-1}{\sqrt{1 - e^{2}}} \left[ \frac{GM_{2}}{v_{\rm W}^{2}} \right]^{2} \frac{\alpha_{\rm W}}{2 a^{2}} \frac{1}{(1 + v^{2})^{3/2}} \dot{M}_{\rm 1W} \:\:{\rm M_{\odot} yr^{-1}}\:\:,
\end{equation}
where $v^{2} = v^{2}_{\rm orb}/v^{2}_{\rm W}$, $v_{\rm orb}^{2} = G M_{\rm b}/a$, subscripts $A$ and $W$ refer 
to accretion and wind quantities respectively and $M_{1}$, $M_{2}$ and $M_{\rm b}$ are the masses of the primary, secondary and whole system 
respectively. The wind speed $v_{\rm W}$ is taken to be proportional to the escape 
velocity $\sqrt{2 \beta_{\rm W} G M_{1}/R_{1}}$. We take $\beta_{\rm W}$ to be 
7 for temperatures above $T_{\rm eff} \sim 21,000\,{\rm K}$ and otherwise $\beta_{\rm W} \sim 2$, to account for the 
bistability jump (Lamers et al. 1995). This is likely an overestimate for cool supergiants.
The parameter $\alpha_{\rm W}$ is taken to be $3/2$. 

Whilst wind AM losses cause some circularization of the orbit, a stronger effect on systems whose periastron distance is 
less than about 30 -- 40 $\rsun$ is tidal synchronization (e.g. Hut 1981). Here again we use the prescriptions given in 
Hurley et al. (2002) to quantify this effect on the orbit. We also use these prescriptions for finding the rate of RLOF.
For finding the onset of RLOF we use the approximation to Roche lobe radius of Eggleton (1983). 

Dynamical-timescale mass transfer is the most common type, and is assumed to lead to rapid spiral-in 
and merging. If there is thermal- or nuclear-timescale mass transfer we assume that accretion is Eddington-limited 
with any excess matter being ejected from the system with specific angular momentum equal to that of the orbital angular momentum of 
the accreting star (King \& Ritter 1999, King et al. 2001a). This may occur for systems with more massive black hole companions.

Comparing these partially-synthetic models to full binary evolution models we find that the fit for non-RLOF systems is 
extremely good, varying by one percent or less throughout the evolution in period and mass. The synthetic-RLOF fit is less good, 
but even so variation is generally less than ten percent. Given that RLOF usually leads rapidly to a merger, this is good enough for our purposes.

\subsection{X-ray luminosity}

The simplest assumption concerning the X-ray luminosity is to assume that some fraction of the gravitational potential 
energy released during the accretion process is converted into X-rays in the 0.2 -- 10 keV band, i.e.
\begin{equation}
L_{\rm X} \sim \epsilon\,\frac{G\,M_{2}}{R_{2}}\,\dot{M}_{2A}  \:\: , 
\end{equation}
where $\epsilon$ is generally assumed to be 0.1 -- 1 (we use 0.5).
In combination with formula 1, this gives an average value of X-ray luminosity from wind accretion throughout the orbit assuming 
uniform winds. This is probably a reasonable estimate for supergiant X-ray binaries and (perhaps) the quiescent phase of 
transient Be X-ray binaries. However the values often quoted for X-ray luminosities of observed systems are {\it maximum} 
values (see e.g. Raguzova \& Popov 2005). For Be X-ray binaries in particular these will not correspond well to theoretical 
Bondi-Hoyle wind accretion values since most Be/X systems are transient (e.g. Negueruela 2005). During quiescence they display 
very low X-ray fluxes as suggested by the wind accretion model used above. Type I outbursts occur mainly at periastron, involve an 
X-ray brightening by a factor $\sim 10$ and are thought to arise from the interaction of the compact object with the 
equatorial decretion disk of the Be star (Laycock et al. 2003). Type II outbursts are rarer occurances in which the X-ray 
emission approaches the Eddington limit for the compact object and, whilst usually starting shortly after periastron passage, 
display no other correlation with orbital parameters. They may arise from the temporary dissipation of the 
Be star disk, allowing a faster rate of accretion onto the compact object than can be achieved through the stellar wind 
(Okazaki \& Negueruela 2001, Negueruela et al. 2001).

As a first approximation to fitting the luminosities observed in X-ray binaries with wind accretion we therefore 
use the following scheme. First we get an idea of whether with standard wind accretion the binary is likely to be 
transient or persistent\footnote{Note that this implicitly assumes the NS accretes mainly via a disk of its own; see e.g. Hayasaki and 
Okazaki (2005). This is rather uncertain.}.
  Following Podsiadlowski, Rappaport \& Han (2002), we use equation A5 of Vrtilek et al. (1990) to 
estimate the outer accretion disk temperature. If it is above 6500 K, we assume the entire disk is ionised and it is 
a persistent system. Otherwise we use the disk instability criterion given by van Paradijs (1996).

For X-ray binaries which are likely to be persistent we use the steady Bondi-Hoyle wind rates as above, taking 
the maximum rather than the average value around an orbit if the system is eccentric. For systems which are likely to be transient, 
following Okazaki \& Negueruela (2001), we distinguish between systems which are likely to undergo type I and type II 
outbursts by eccentricity. For $e \gtrsim 0.6$ we assume type I outbursts only and for $e \lesssim 0.2$ type II outbursts only.
For systems with intermediate eccentricities the expected outburst type is rather sensitive to the exact circumstances of the binary, 
and both sorts of outbursts may occur. The $e = 0.34$ transient Be/X binary 4U 0115+63 (Negueruela \& Okazaki 2001), which 
displayed a series of type I outbursts in 1996 after previously being known as a type II outburst source, is an example of such a system.

As noted by Van Bever \& Vanbeveren (2000), Type II outbursts are relatively poorly-understood; we assume the accreting 
star will do so at an Eddington-limited rate. There do exist circumstances which may lead to apparent super-Eddington accretion, 
such as X-ray beaming (King et al. 2001b). 
For type I outbursts we use the Bondi-Hoyle wind accretion formula as for supergiant systems, but assume the maximum 
luminosity comes when the compact object passes through the equatorial wind of the Be star.
One explanation for the Be star disk is that the fast rotation of the star leads to latitude-dependent wind velocities 
(Lamers \& Pauldrach 1991). Detailed calculations of this effect for the (similar) B[e] stars (Cur{\'e}, Rial \& Cidale 2005; 
Pelupessy et al. 2000) suggest that the equatorial wind velocity in this case may be reduced by a factor 3 -- 6 for 
rotational velocities typical of Be stars, whilst the equatorial mass flux is little-changed. We therefore alter our 
wind velocities accordingly. It should be noted that the X-ray luminosity is very sensitive to the wind velocity through its 
effect on the amount of matter accreted. Furthermore, these quantities may well be subject to short-term fluctuations 
which it is difficult to account for (Podsiadlowski et al. 2002) and the accretion rate itself may depend strongly on local 
turbulence (Krumholz, McKee \& Klein 2005). Therefore in general the X-ray luminosities given here should be regarded as approximate.

There are also a number of further factors which may keep systems from being X-ray bright. In the case of stellar wind accretion 
on to a BH, in order for significant X-ray emission to be produced one requires the formation of a disk (e.g Iben 
et al. 1995). However in the wind accretion case it is difficult to form an accretion disk at all due to the typically low specific angular 
momentum of the captured matter (Illarionov \& Sunyaev 1975). Unless the specific angular momentum of the accreting 
matter is greater than that of a particle 
in the innermost stable orbit around the BH, no disk is formed. This may be used to derive an orbital period limit above which 
a BH-containing system will be X-ray dark,

\begin{equation}
P_{\rm orb} > 4.8\,(M_{\rm BH}/\msun)\,(v_{wind}/10^{4} {\rm km\,s^{-1}})^{-4}\,\delta^{2}\:\: \rm{hr} ,
\end{equation}

where $\delta$ is a dimensionless parameter of order 1 (for details see Ergma \& Yungelson 1998). Massive stars such as are observed in 
HMXBs typically have high wind velocities (100 -- 3500 ${\rm km\,s^{-1}}$ for O and B stars, 1000 -- 4000 ${\rm km\,s^{-1}}$ for 
WR stars). The limiting period below which a BH system is X-ray bright may therefore be as small as a few hours. We implement this 
limit in our simulations. There is also the possibility of a proportion of rapidly-rotating NS binaries which are X-ray dark due to the 
propeller effect (see section 5.1).

For systems which accrete via RLOF we use equation 2. However as noted above accretion and luminosity are capped at the Eddington limit, 
\begin{equation}
L_{\rm edd} = 2.5 \times 10^{38} \frac{M_{compact}}{M_{\odot}} / (1 + X)\:\: \rm{erg\,s^{-1}},
\end{equation}
where $X$ is the hydrogen abundance by number of the accreting matter. This may still produce systems with rather high 
X-ray luminosities if the compact object mass is large and the accretion stream is H-poor.

As has been noted previously, systems which accrete at periastron will be X-ray dark for most of their orbits, 
whereas compact objects accreting via winds from a companion in a close low-eccentricity orbit or RLOF will be
bright all the time. This suggests the latter type of system will be significantly over-represented in 
observed populations. However, since the completeness of the transient sample is related to the orbital period and 
the times that an area has been observed, attempting to reproduce the distribution of maximum luminosities (as opposed 
to the total overall luminosity) by weighting by duty cycle will also not be accurate. Similarly the currently observed 
maximum luminosity of some systems may not be the maximum thoretically-attainable luminosity, most notably if thus far 
they have only been observed in quiescence. Currently we do not apply any 
weighting to account for either of these effects, which will lead to under-representation of long-period transients and 
over-representation of high-luminosity systems with respect to the observed population. 

\section{Galactic HMXBs}
The Milky Way sample of HMXBs includes those systems which are nearest to us and for which the orbital parameters
are best-determined. In particular, many more Galactic than extragalactic systems have known periods and 
eccentricities, and more have known individual object masses (table 1).
It is appropriate, therefore, to test that our models are reasonable against this sample
before drawing conclusions about the metallicity dependence.
Recent HMXB catalogues are Liu, van Paradijs \& van den Heuvel (2000) and Raguzova \& 
Popov (2005; includes only Be/X-ray binaries). Unless otherwise stated, all observational results shown 
are obtained from these catalogues and references within. In figure 1 we show this observational sample against the 
theoretical distributions we have obtained, for conservative RLOF. Figure 2 shows the corresponding distributions for 
non-conservative RLOF. Note that we omit the extremely short-period system Cyg X--3 from the plots, as there is some doubt about its HMXB nature
(Vanbeveren, de Loore \& van Rensbergen (1998); see Lommen et al. (2005) for a comprehensive discussion of this system). 
Whilst we can fit Cyg X--3 with our models, it is only as a rare and transient stage of a system about to merge.

\begin{figure*}
\vbox to205mm{\vfil
\psfig{figure=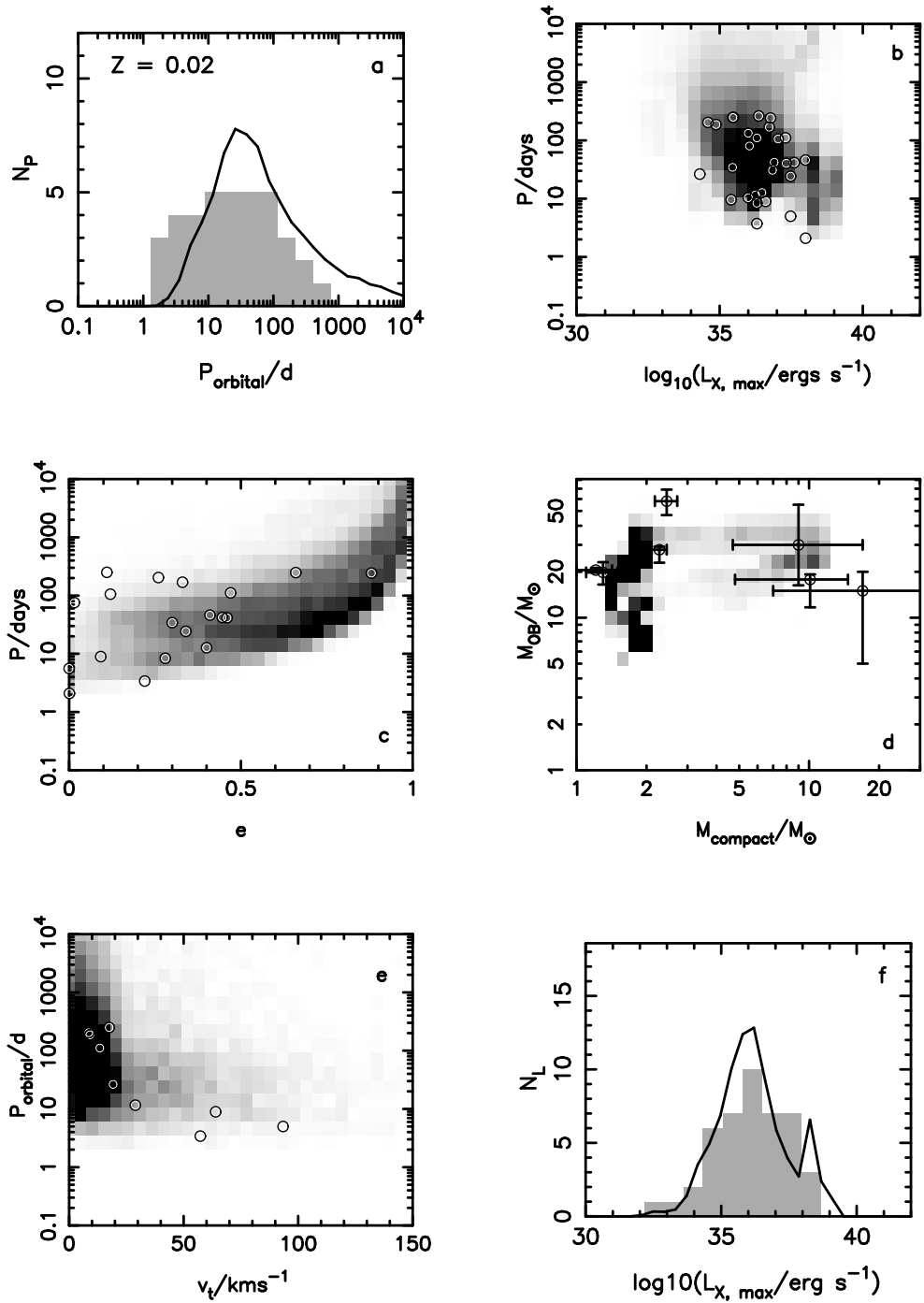,angle=0,width=130mm}
\caption{Theoretical properties of the simulated set of luminous HMXBs versus observation, for 
solar metallicity and conservative mass transfer. In panel a the normalised period distribution is shown 
against the observed period distribution of Galactic HMXBs (Liu et al. 2000, Raguzova \& Popov 2005).
In panels b, c, d and e the observed (points) and theoretical (greyscale) maximum luminosity, eccentricity, 
component mass and transverse velocity distributions are shown. Note that the subset of the HMXB 
population in each set of observed values is different. The maximum luminosity and period distributions are 
the most complete. Very low-luminosity systems (below $10^{32} {\rm ergs\: s^{-1}}$) are omitted from all plots 
except those detailing luminosity distributions.}
\vfil}
\label{fig1}
\end{figure*}

Nearly all systems in both mass-transfer scenarios have periods above one day. Considering a typical NS-Be binary with masses 
$1.4 \msun + 10 \msun$, such a short period would result in a separation of only 9 solar radii and the Be star would 
overflow its Roche lobe whilst still near the main sequence. Given that OB stars in close X-ray binaries 
have already lived at least long enough for their companions to evolve to the post-SN stage, their radii will be 
already rather larger than ZAMS radii and so RLOF and a rapid merger is likely.
Hence only atypical HMXB systems are likely to exist for any length of time 
with periods shorter than a day, and it is unsurprising that even including the effects of orbital evolution we find 
few systems below this limit. The drop-off in number of systems with periods over about 100 days is partly due to the input 
progenitor population period distribution, but is also due to our implementation of a luminosity limit (currently 
$10^{32}\,{\rm ergs\:s^{-1}}$) below which we assume a system is unlikely to be observable even in relatively deep surveys.
Longer-period systems, unless their eccentricity is also very high, frequently fall below this limit. Given that it becomes 
increasingly unlikely that a system has been observed in outburst as the period becomes larger, the offset towards high periods 
of the model period distribution compared with the observed distribution seems reasonable.

\begin{figure*}
\vbox to188mm{\vfil
\psfig{figure=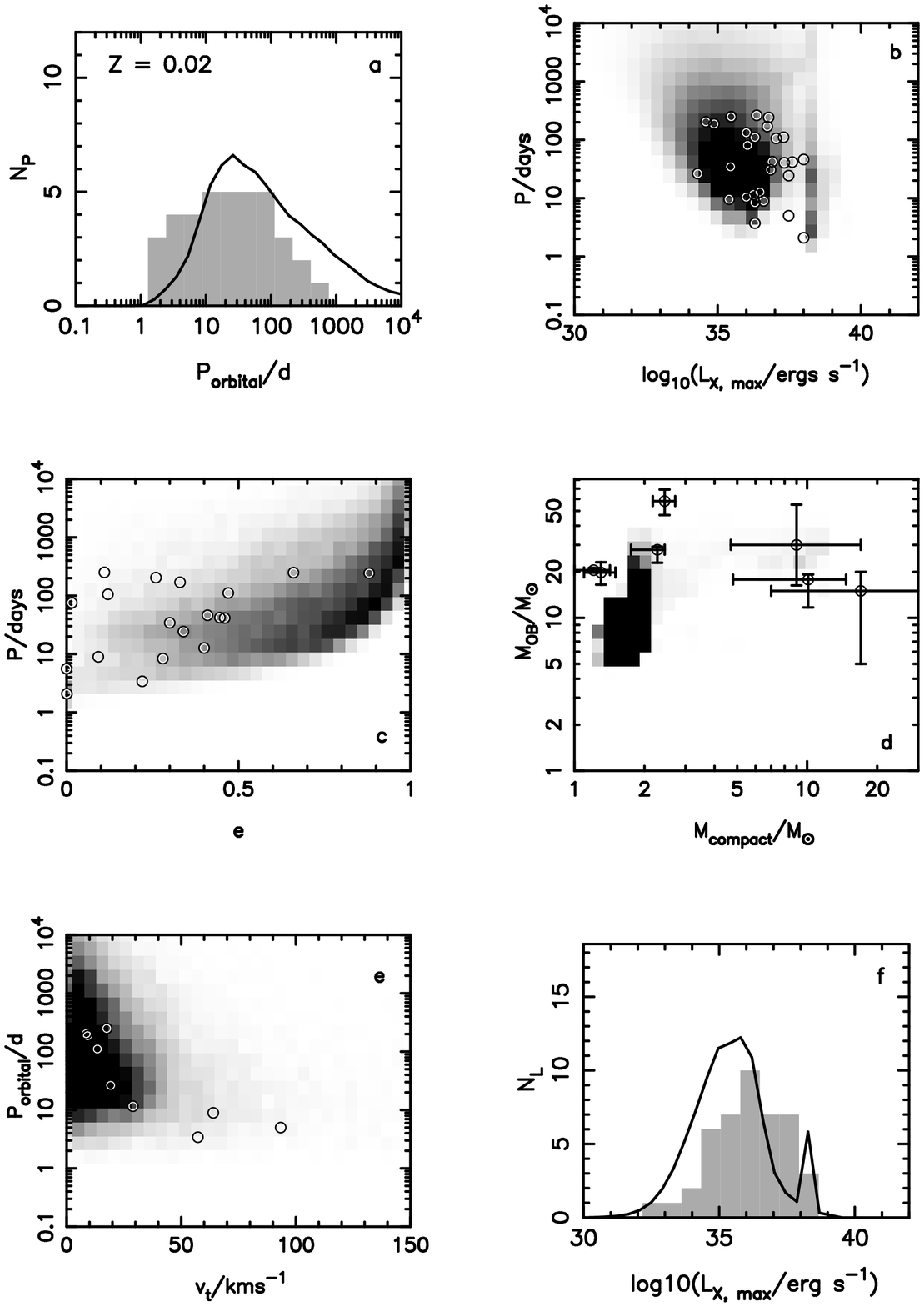,angle=0,width=130mm}
\caption{As figure 1, but for non-conservative mass transfer at solar metallicity.}
\vfil}
\label{fig2}
\end{figure*}

The bounds of the eccentricity distribution are also determined in part by external factors (see e.g. Pfahl et al. 2002, 
with which our eccentricity distributions agree well), meaning that the main effect of following the time evolution of 
systems is a shift towards low- and zero-eccentricity systems which have circularised. The lower period bound on 
eccentricity reflects the periastron distance at which RLOF occurs. The drop-off at high period is again the 
effect of the input population combined with the luminosity limit. In particular, there are relatively few long-period 
systems because pre-SN long-period systems are less strongly bound and so are more likely to be split or given a 
large eccentricity by any SN kick. It should be noted, both here and for the velocity distribution (panel e) that 
since we have not weighted the observability of systems by duty cycle, transient systems (generally long period 
and/or high eccentricity)  are somewhat over-represented at the expense of persistent 
systems. The systems residing in the high-velocity tail tend to be persistent (as they are in real life; see van den Heuvel 
et al. 2000), so in the distribution of systems currently observable (i.e. persistent systems plus systems which have had an 
outburst since large-scale X-ray observations began) the high-velocity tail would contain a greater share of the population.

The mass distribution of components (panel d) is bounded on the upper end of companion mass and the lower end of 
compact object mass by the progenitor population, as significant wind or RLOF mass loss is likely and mass gain unlikely for companion stars. 
Systems containing a very massive companion with a black hole are 
present in the progenitor population but not well-represented in the actual population 
since they require for formation two initially massive stars whose lifetimes will both be only a few $10^{6}\,$yr. Thus after the 
first SN the second star will be very short-lived. It is notable that we find some bimodality in the distribution of compact object masses, 
with relatively few having 
masses between 2 and $6 \msun$. This occurs despite the assumption of a smooth function for mass lost in the SN in relation 
to core mass over the NS-BH transition (see D05; the maximum NS mass is assumed to be $2.2 \msun$ in this paper). The existence of such 
a gap in real life has been postulated for low-mass X-ray binaries by 
Bailyn et al. 1998, and is certainly not ruled out by the limited sample of high mass X-ray binary masses available 
(see table 1). Fryer \& Kalogera (2001), who find a theoretically continuous distribution of black hole masses, suggest that 
the apparent gap may be a selection effect.

The gap in our models arises from the combination of several effects. The period limit we have imposed on 
black hole binaries (equation 3) favours more massive black holes. However, the gap is also present in the progenitor 
distribution to an extent. This is likely a consequence of the SN mass loss distribution that we assume. Although the 
remnant mass distribution with initial core mass is smooth over the NS--BH transition, the proportion of a system's mass which is lost in the SN 
peaks for remnant masses of around $4 \msun$. In combination with relatively large SN kicks, this leads to a 
greater proportion of systems which will form black holes of around this mass being split by the first SN. In this 
picture, the total true underlying distribution of compact objects is indeed smooth -- however, those around the 
black hole `mass gap' are more likely to be single and hence undetectable. 
It is also apparent from the mass distribution that at least some conservative mass transfer systems are required 
to match observed masses. 

It should also be noted that the distributions shown are normalised to the observed Galactic population. However, it is relatively 
simple to make an estimate of whether the absolute rate is reasonable. For the region within $2.5 {\rm kpc}$ of the sun, 
van Oijen (1989) gives an estimate of the total number of HMXBs to be 50 Be systems and 3 supergiant systems, for an O star population of 960 -- 
1850\footnote{Note that the local massive star population derives from two different sources: the Galactic disk and an extra population 
from the Gould belt. Therefore massive stars as a whole may be relatively overabundant in the solar neighborhood.}.
For constant star formation, our solar metallicity models over the full range of input parameters suggest a population 
of between 27 and 112 O stars per HMXB, i.e. using the estimates of O star population above, around 10 -- 70 HMXBs in 
total within $2.5 {\rm kpc}$ of the Sun. Values around the lower end of this range are obtained using conservative mass transfer models; 
the upper end of the range is from non-conservative models.
Since the solar neighborhood (and quite possibly also the Sun, e.g. Apslund et 
al. 2005) is at a lower metallicity than the value commonly used as `solar', it should also be noted that the corresponding range for 
our next metallicity down, ${\rm Z = 0.01}$, is similar, extending upwards to slightly higher values (the HMXB population increases 
but so does the O star population). These estimates seem reasonable.

Since we have found a reasonable match to observed Galactic populations, it seems reasonable to now extend our models to other 
metallicities in order to quantify the metallicity effect on HMXB populations. 

\section{Metallicity}

The primary sources for observations of binary systems at metallicities lower than solar are the Magellanic Clouds.
The LMC, at metallicity approximately half solar, has a HMXB population which is not strongly dissimilar from the Galactic one 
(Shtykovskiy \& Gilfanov 2005a). The SMC has metallicity around one-fifth solar, and has a very large Be/X-ray population probably 
related to recent star formation (Majid et al. 2004; Haberl \& Pietsch 2004). 

\begin{figure*}
\vbox to190mm{\vfil
\psfig{figure=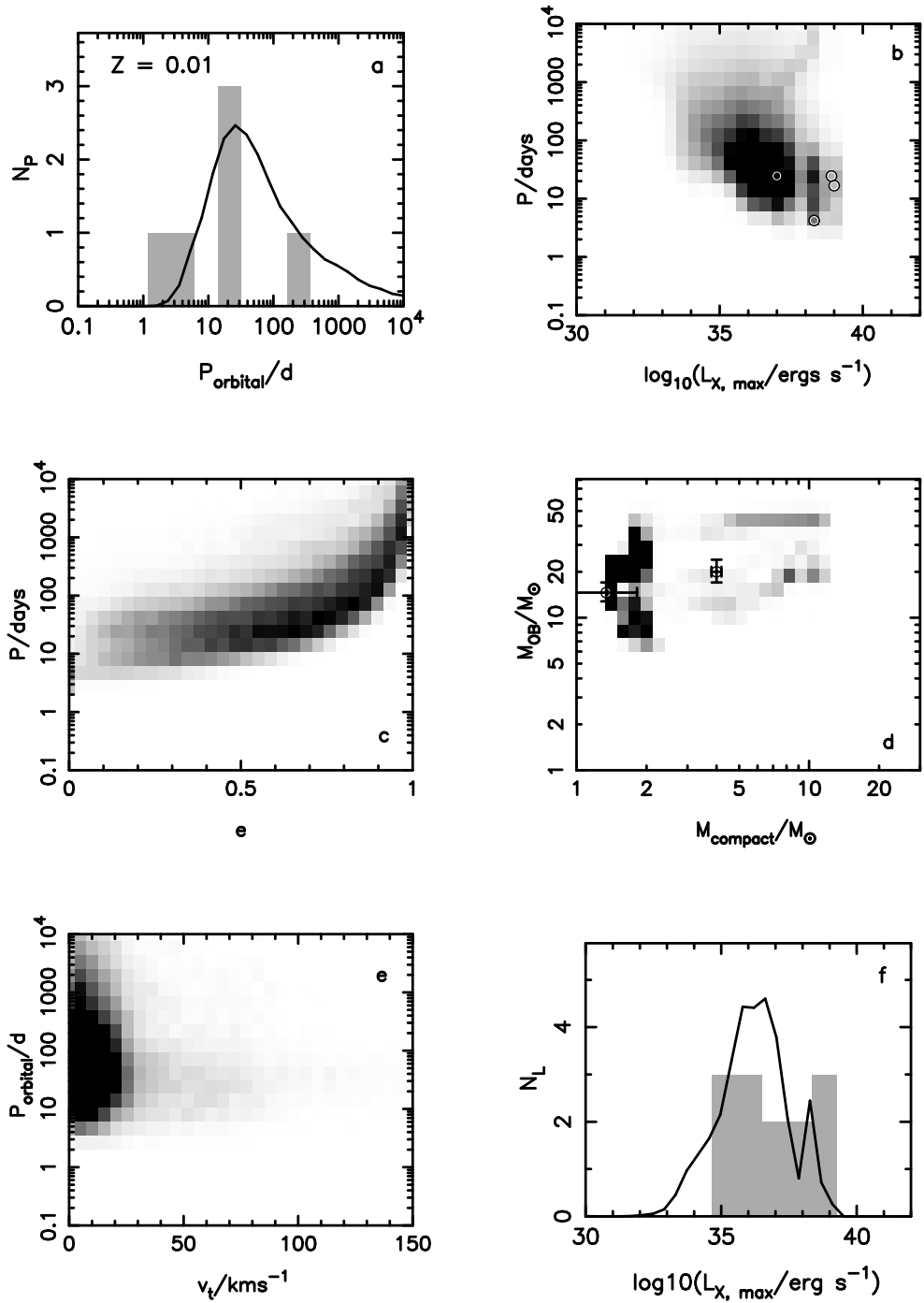,angle=0,width=130mm}
\caption{As figure 1, but for half solar metallicity in comparison with LMC observations.}
\vfil}
\label{fig3}
\end{figure*}

As has been noted in the introduction, mass loss from line-driven winds is less strong at low metallicity. Between 
solar and SMC metallicity one would expect a factor 2 -- 4 decrease in mass-loss rate between comparable O stars. This affects 
the evolution of both components. With weaker winds, stars arrive at the SN stage with higher masses. Final core masses are 
also generally greater at lower metallicity for equivalent models (e.g. Heger et al. 2003), although this has the potential to be 
complicated by a number of other effects varying from star to star, in particular rotation. Higher core masses in turn are likely to result in 
higher remnant masses, i.e. more BH XRBs at the expense of the NS XRB population.
Similarly the companion star's mass will be lost less quickly, so the average companion mass will be higher. 
Less matter is lost in the wind but there is a slight trend towards smaller wind velocities at lower metallicity 
(Prinja \& Crowther 1998) so the change in amount of matter accreted with metallicity is not straightforward. We note 
from the theoretical results of Kudritzki, Pauldrach \& Puls (1987) that the wind velocity metallicity dependence may be 
approximated roughly as $v_{\infty} \sim Z^{0.14}$, and implement this in our low metallicity simulations.
The wind also takes away angular momentum, so one would expect closer orbits on average for 
lower-metallicity systems, which in turn will affect the proportion of systems which are split by SNe. Assuming 
SN kicks follow a similar distribution 
to that at solar metallicity, the compact object-containing binary population at low metallicity should be greater than that at 
solar metallicity.
Of course, since the mechanism of SN kicks is not yet certain, it is possible that they depend on metallicity as well, if not 
directly then by the effect of metallicity on evolutionary parameters which might make a difference, such as rotation 
(Podsiadlowski et al. 2004). We currently 
assume that this is not the case.
Disregarding the question of whether they are observable or not, then, we expect the population of potential X-ray binaries to increase as 
metallicity decreases, and that these systems should be on average closer and more massive. In particular, we expect an increase 
in BH binaries, in agreement with Belczynski et al. (2004a) and Hurley, Tout \& Pols (2002). 

\begin{figure*}
\vbox to192mm{\vfil
\psfig{figure=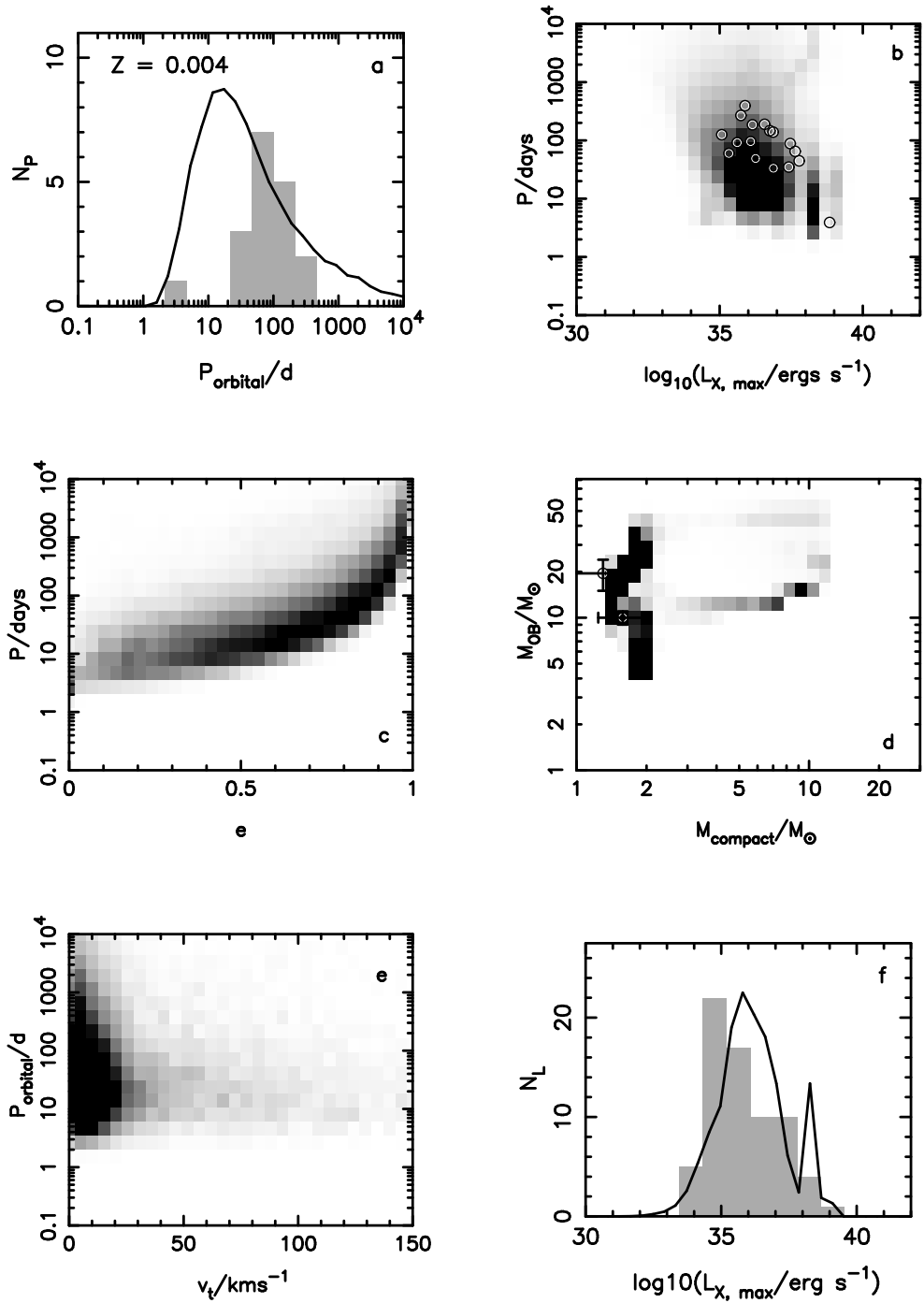,angle=0,width=130mm}
\caption{As figure 1, but for $Z = 0.004$ in comparison with SMC observations. Luminosity observations are from 
Shtykovskiy \& Gilfanov (2005b) and Raguzova \& Popov (2005) with the former values used when the two 
catalogues overlap.}
\vfil}
\label{fig4}
\end{figure*}

For our LMC-appropriate full sets of models (figure 3) we seem to get a reasonable 
fit using the same parameters as for the Galactic models, although given the small number of observed HMXBs in the LMC  
the significance of this result is low; it is also notable here again that short-period persistent systems are more readily observed than 
long-period transients.

For our SMC models, however, we have notably different results. In particular, it is impossible to fit the current period 
distribution with a constant SFR (figure 4). This should come as no surprise, since most of the SMC distribution is thought to result from 
a recent intense burst of star formation (e.g. Haberl \& Sasaki 2000). A more suitable test is whether a starburst-type population 
can provide a fit to the period distribution at any age. In particular, the observations of Harris \& Zaritsky (2004) suggest that 
there were bursts of star formation in the SMC 60 Myr and 400 Myr ago. From our models we would expect an increase of around a factor 
of three (potentially more if accretion is not Eddington-limited) in the possible SMC luminous HMXB population for the same amount of 
star formation when compared to our Galaxy. 
However, the SMC is in fact overabundant in HMXBs for its size when compared to the Galaxy by as much as a factor of 50. 
This also suggests active star formation. Shtykovskiy \& Gilfanov (2005b) find different star formation rates are indicated by different 
methods, but if the far-infrared, H$\alpha$- and ultraviolet-based estimators are correct then HMXBs are around a factor of 10 overabundant 
when the SFR is accounted for. The differences in distribution are unlikely to be due to any 
plausible selection effect, since it is the shorter-period, persistent systems -- supposedly more observable -- which are missing.

Over the lifetime of a coevally-formed population of stars, the HMXB distribution is likely to change significantly. Close 
supergiant X-ray binaries are strongly represented in the young population, but die out relatively early. The older bright 
HMXB population is characterised by Be companions, wider, eccentric orbits and compact objects that are almost exclusively NSs.
Figure 5 shows the distribution obtained in period and luminosity during different epochs for SMC metallicity, in 
comparison with the observed distributions. It is apparent that it is hard to reconcile both the luminosity and period distributions. 
Whilst both compact object and companion masses measured for the SMC (table 1) fall into the narrow range of masses for 
old bright HMXB systems, there is still an excess of theoretical short-period HMXBs when compared with the observed, primarily long-period,
SMC systems. It is also notable from Fig. 5 that the HMXB population declines with time. Older HMXB populations have longer peak periods, 
and so are closer to the observations. However, they are also significantly less numerous, requiring even more intense star formation 
to have taken place.

\subsection{The propeller effect}

Young NSs in HMXBs are likely to have significant magnetic fields. In systems which have a rotating NS as their 
compact component, this implies a transition at some radius $R_{\rm m}$ between disk-like accretion and magnetospheric
accretion co-rotating with the NS (Lamb, Pethick \& Pines 1973). $R_{\rm m}$ is greater at smaller values of 
the mass accretion rate and hence also at smaller values of the X-ray luminosity.
If the mass accretion rate is low, $R_{\rm m}$ may be large enough that co-rotating matter at $R_{\rm m}$ would exceed the
Keplerian angular velocity. This inhibits accretion; the transferred matter is effectively expelled from the system by 
centrifugal force. Whilst this is occurring one would expect very low X-ray emission. Such an effect has been 
suggested as the explanation for the cessation of X-ray emission in GX 1+4 (Cui \& Smith 2004).
However, the systems affected by the propeller effect should form the low-luminosity tail of the HMXB population. 
Whilst Shtykovskiy \& Gilfanov (2005a) find that the propeller effect provides a fitting explanation of the lack of low-luminosity 
X-ray binaries in the LMC, in the SMC by contrast there is an apparent excess of low-luminosity binaries in comparison to 
the expected constant-SFR population. If instead one assumes the later end of the old HMXB population, as suggested above, it is possible to 
account for both the luminosty and period distributions. However doing so 
requires that the currently-visible HMXB population in the SMC represents only a small fraction of the population which 
would be visible in the absence of the propeller effect. Given that the SMC is already significantly overabundant in 
HMXBs, this seems a very non-optimal (although potentially viable, if the HMXB population increases with decreasing metallicity and there 
is also vigorous star formation) solution. 

\begin{figure}
\vbox to136mm{\vfil
\psfig{figure=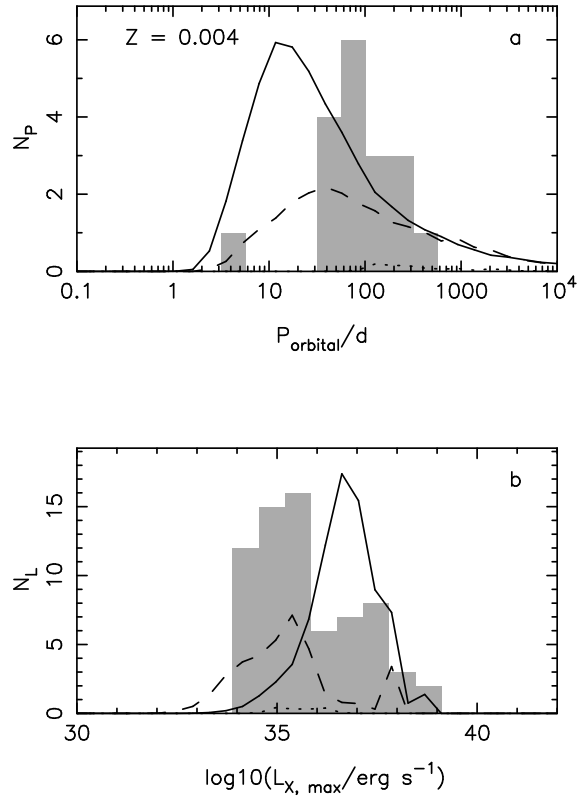,angle=0,width=75mm}
\caption{Period and luminosity histograms as in previous figures for young ($\sim$ 12 Myr, solid line), middle-aged 
($\sim$ 36 Myr, dashed line) and old ($\sim$ 124 Myr, dotted line) HMXB populations in our models. The relative sizes 
of the two populations are preserved, i.e. the young HMXB population is much more numerous. The old population is 
almost entirely composed of Be star/NS binaries. It is apparent that, whilst the observed luminosity distribution can 
be reproduced by a judicious combination of starbursts, the period distribution is representative of a primarily old
population. Since the old population is less numerous, this requires a very high historical SFR.}
\vfil}
\label{fig5}
\end{figure}

One must also address the question of why the propeller effect would occur to different extents in the Galaxy, LMC and SMC. 
Theoretically it has been suspected for some time that rotation rates are faster at lower metallicity (e.g. Maeder \& 
Meynet 2001) due to the decreased angular momentum loss from winds, but observationally there is some evidence that 
this may not be the case (Penny et al. 2004). One would also expect a larger fraction of the population to contain BHs at 
low metallicity, again due to lower mass-loss rates; these systems should be unaffected by the propeller effect, although 
since in the old HMXB population scenario the surviving luminous systems contain mostly NSs close to $1.4 \msun$, this will 
only make a difference for young populations. Shorter-period systems are also more likely to have undergone a common envelope 
phase (currently rather poorly-understood) in their past evolution, so it is possible the shorter-period end of the old HMXB 
population is theoretically over-represented because our assumptions about the evolution of such systems are inaccurate. Since 
the old HMXB population is smaller than the young HMXB population, this would have only a weak effect on the constant-SFR populations 
we have compared the Milky Way and LMC against, but a large effect on the SMC where the old population dominates.

Another possibility is that the shorter-period systems exist, but are in quiescent states which are not sufficiently 
luminous for us to observe. This implies type II outbursts with 
the duty cycle parameter low. Since the missing SMC HMXBs are systems with periods below 10 -- 20 days it is 
possible that they have tidally circularised, so this situation is reasonable. However it would also suggest again that there are 
currently even more HMXBs in the SMC than the very large population we currently know.

One further way of matching the period and luminosity distributions simultaneously would be to have a low-eccentricity population of moderately 
wide Be star X-ray binaries. This could be achieved if some proportion of such systems have small SN kicks. Since large kicks split the 
binary in 70 -- 80 \% of cases whereas small-kick systems are much more likely to stay together, this is potential evidence 
for a fraction of systems, at least at low metallicity, having small kicks (Pfahl et al. 2001, Podsiadlowski et al. 2004). 
Even if not very many systems follow this route, the much increased liklihood of their remaining a binary after the first SN
could both provide a much larger HMXB population than we would expect from models in which all SNe have large kicks and also 
strongly skew the HMXB parameter distribution. However as noted in D05, unless the fraction of small-kick systems is very small 
there is a problem with creating enough runaway stars. If this mechanism is active in the SMC population we would expect a 
corresponding decrease in the frequency of runaways.

\subsection{Overall metallicity trend}

In figure 6 we show the overall metallicity number trend for constant-SFR luminous HMXB systems. As expected, there is generally an increase 
in HMXB population with decreasing metallicity. Perhaps surprisingly, this applies both to the NS and BH systems, however. 
An increase in HMXB population of around a factor of 3 between Solar and SMC metallicity is what we would expect from looking at our
progenitor population and this is generally confirmed in the populations of luminous systems produced. If accretion is not Eddington-limited and 
if the the condition that most BH binaries are not X-ray luminous (equation 3) is relaxed somewhat (e.g. if the wind velocities are 
lower than has been assumed) then this increase can be up to a factor of 10.

\begin{figure}
\vbox to125mm{\vfil
\psfig{figure=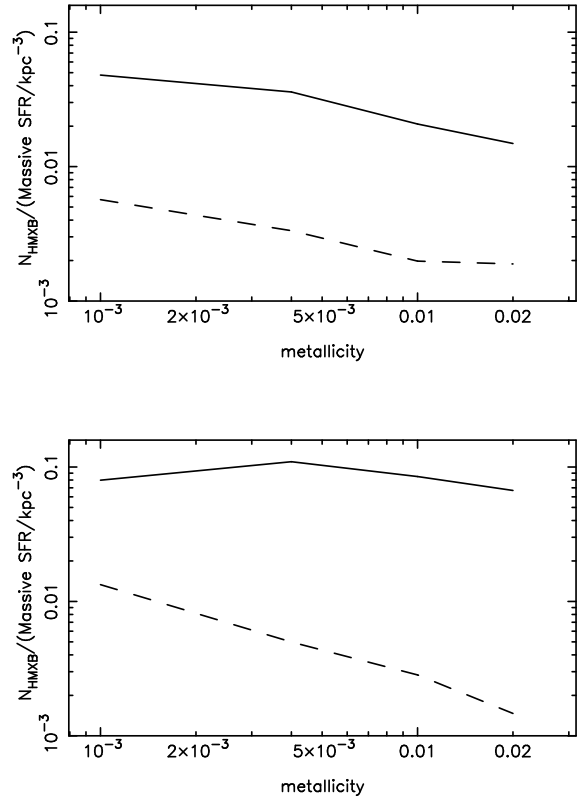,angle=0,width=75mm}
\caption{Luminous HMXB population per ${\rm kpc^{3}}$, per unit of massive ($> 10 \msun$) SFR in ${\rm stars\,  Myr^{-1}\, kpc^{-3}}$, 
for conservative (upper panel) and 
non-conservative (lower panel) mass transfer. Note that the population is log-scaled. The solid/dashed lines indicate binaries in which 
the compact object is a NS/BH.}
\vfil}
\label{fig6}
\end{figure}

Belczynski et al. (2004a) find an increase of around a factor of 4 in the BH binary population between ${\rm Z = 0.02}$ and $0.001$.
Given that this includes all BH binaries -- not just the fraction that are luminous -- this seems reasonably consistent with 
our results. Between ${\rm Z = 0.02}$ and $0.0001$, Hurley et al. (2002) find an increase in populations of a factor of 2. 
It is possible that, given the wider metallicity range, this represents a slight downturn in HMXB populations once the metallicity becomes 
very low. We see a downturn in non-conservative BH systems at low metallicities -- such an effect could 
increase at very low Z due to the larger proportion of SN mass loss as opposed to wind mass loss contributing 
to a greater splitting rate of binaries, and/or smaller wind mass loss leading to a greater proportion of remnants becoming 
BHs which may then not be X-ray luminous.

Other consistent effects in the model population which we observe with metallicity are generally relatively small. With decreasing 
metallicity, there is a small decrease in the peak period and the period distribution extends down to slightly shorter periods, as
the smaller radii of lower-metallicity stars help systems to avoid RLOF for longer. There is also a slight trend towards higher-eccentricity 
systems, and towards a greater proportion of Eddington-luminosity systems. However in general our models suggest it is safe to assume 
the general parameter distribution of HMXB systems of similar ages is roughly the same over this range of metallicities.
 
\section{Discussion}

From previous sections, it appears that theoretical and observational agreement is reasonable for the Galaxy and LMC, but there remain 
some questions about the HMXB population of the SMC; it is hard to produce such an overabundence of long-period systems without 
an extremely intense starburst. This model failure, in the light of the relative success with the other populations, suggests either 
that there may be other effects coming into play for the SMC population which have not been considered here, perhaps related more specifically 
to the SMC environment (e.g. star formation in regions of high turbulence) or else that in general there is an extra effect which we 
have not considered which becomes more important once the metallicity is below some threshold value. However, in that case all starburst
environments with populations of comparable age to the young SMC population and similar or lower metallicity should be similarly highly 
overabundant in HMXBs. One such effect could be, for example, the different evolutionary paths followed by very rapidly-rotating stars, 
if there are more such stars at low metallicity (Woosley \& Heger 2006). The easiest way to create much larger HMXB populations is to have 
smaller SN kicks, because then a larger proportion of systems survive the first supernova. Thus the failure of our model for 
the SMC could be taken as indirect evidence for smaller kicks at low metallicity, again possibly related to the behaviour of rotation with 
metallicity.
 
There remain a few other interesting issues and conclusions which may be drawn from our simulations.
It is notable, particularly in the light of D05 in which non-conservative RLOF produced a 
better match with the observed population, that some systems are difficult to fit with non-conservative mass transfer.
This has previously been noted by Wellstein \& Langer (1999) in the case of GX 301-2, 
although there is also significant evidence that non-conservative mass transfer must take case 
in some cases (van den Heuvel et al. 2000). The problem lies 
in producing a sufficient number of neutron stars close to $1.4 \msun$ which have $\sim 20 \msun$ or greater companions such as Cen X--3 and 
QV Nor. We would expect therefore if these mass measurements are correct that at least some systems have (quasi-)conservative 
mass transfer. The better fit of 
conservative systems to the X-ray binary population than to the runaway population may 
in fact be entirely consistent, however. In real life, it seems likely that some RLOF is conservative and some 
is not, depending on, for instance, the rotation rates and mass ratio involved 
(Langer, Wellstein \& Petrovic 2003; Petrovic, Langer \& van der Hucht 2005). It might be considered that 
those systems which are conservative 
will end up at the pre-SN stage with more massive secondaries (albeit in generally wider orbits due to the loss 
of angular momentum in non-conservative mass transfer), and may hence be less likely to be split by the SN. However 
in D05 rates of binary splitting did not differ significantly between the two scenarios. A conservative system which 
is not split is also likely to have greater accretion rates than its non-conservative equivalent since the more massive 
companion star will have greater mass-loss rates. This may make conservative systems brighter in X-rays,
suggesting a significant observational bias in favour of their detection.

A further consequence of accretion in compact binaries is that, in some cases, a neutron star may accrete enough mass that 
it exceeds the maximum NS mass. We have assumed in our simulations that this situation results in the `quiet' formation of 
a black hole. However, it has been recently suggested that such a scenario might, analagously to type Ia SNe, power 
a short gamma-ray burst (McFadyen, Ramirez-Ruiz \& Zhang 2005). If this is the case, we would expect roughly one in $10^{5}$ 
stars over $10 \msun$ to undergo this fate if accretion is not Eddington-limited, or as few as one in $10^{7}$ if it is.
How many of these would be observable would, of course, depend on the degree of beaming involved and the intrinsic brightness 
of such an event. However it is notable that, certainly in the Eddington-limited case, the event rate of such transitions is 
significantly lower than we would expect for long GRBs, assuming  1\% of SN Ib/c produce an observable long burst (Granot \& Loeb 2003).

Finally, though we have limited our models to the Eddington luminosity, it is also interesting to note their behaviour when not so limited (Rappaport et al. 2005).  
In particular, we find that $L_{X}$ values as great as $10^{41} {\rm erg\,s^{-1}}$ may be (briefly) attained in some cases. A system observed with 
this luminosity would be a particularly bright ULX. The association with star-forming regions, 
particularly in the Cartwheel galaxy, implies that at least some ULXs are relatively young objects and probably an extension of the 
high-luminosity HMXB distribution (King 2004). Some LMXBs have periods of apparently super-Eddington luminosity (Grimm et al. 2002) and mechanisms have been 
proposed by which a HMXB containing a stellar mass BH can exceed the Eddington limit, such as beaming (King et al. 2001b) and 
photon-bubble instabilities in the disk (Ruskowski \& Begelman 2003). If these mechanisms are responsible for some ULXs then from our models we 
would expect this ULX population to peak a little before $10^{7}$ years after a 
burst of star formation, and for each individual ULX to be a relatively short-lived phenomenon. The potential ULX population increases 
sharply (faster than the HMXB population) with decreasing metallicity, although it is also confined to a shorter time period.

\section*{Acknowledgements} 

LMD gratefully acknowledges support from the Leicester PPARC rolling grant for
theoretical astrophysics, and Andrew King and Martin Beer for proof-reading.

{\vspace{0.5cm}\small\noindent This paper
has been typeset from a \TeX / \LaTeX\ file prepared by the author.}

\label{lastpage}

\end{document}